\documentclass[twocolumn,aps,prl,showpacs,floatfix,superscriptaddress]{revtex4}
\usepackage{amsmath}
\usepackage{amsfonts}
\usepackage{amssymb}
\usepackage{graphicx}
\usepackage{epstopdf}
\usepackage{subfigure}
\usepackage{verbatim}
\usepackage{color}

\newcommand{\Rl}{R_\lambda}

\newcommand{\fw}{\sigma_k}
\newcommand{\ooe}{{1/e}}

\newcommand{\rms}{rms }

\newcommand{\transverse}{horizontal }
 
\newcommand{\axial}{axial }

\newcommand{\avar}{\langle a_i^2 \rangle}

\newcommand{\pdf}{PDF\ }

\begin{document}

\title{
 Fluid Acceleration in the Bulk Turbulence of Dilute Polymer Solutions
}

\author{Alice M. Crawford} 
\altaffiliation{Current address: 1204 Sarah Drive, Silver Spring MD 20904, USA}
\affiliation{Laboratory of Atomic and Solid State Physics, Cornell University, Ithaca, NY 14853, USA}

\author{Nicolas Mordant} 
\altaffiliation{Current address: Laboratoire de Physique Statistique, Ecole Normale Sup{\'e}rieure \& CNRS, 24 rue Lhomond, 75005 Paris, France}
\affiliation{Laboratory of Atomic and Solid State Physics, Cornell University, Ithaca, NY 14853, USA}

\author{Haitao Xu}
\affiliation{International Collaboration for Turbulence Research}
\affiliation{Max Planck Institute for Dynamics and Self-Organization, 37077 G{\"o}ttingen, Germany}

\author{Eberhard Bodenschatz}
\email{eberhard.bodenschatz@ds.mpg.de}
\affiliation{Laboratory of Atomic and Solid State Physics, Cornell University, Ithaca, NY 14853, USA}
\affiliation{International Collaboration for Turbulence Research}
\affiliation{Max Planck Institute for Dynamics and Self-Organization, 37077 G{\"o}ttingen, Germany}
% \affiliation{Sibley School of Mechanical and Aerospace Engineering, Cornell University, Ithaca, NY 14853, USA}

\date{\today}

\begin{abstract}
We report experimental measurements of Lagrangian accelerations in the bulk of intense turbulent flows of dilute polymer solutions by following tracer particles with a high-speed optical tracking system. We observed a significant decrease in the acceleration variance in dilute polymer solutions. The shape of the normalized acceleration probability density functions, however, remain the same as in Newtonian water flows. 
We also observed an increase in the turbulent Lagrangian acceleration autocorrelation time in these dilute polymer solutions.
\end{abstract}


\maketitle

It has been known for many years that the addition of a very small amount of long-chain polymers to a fluid can have a dramatic effect on the turbulence near solid boundaries and result in  significantly reduced flow resistance~\cite{toms:1948}. This ``drag-reduction'' phenomenon has been the subject of extensive studies~\cite{lumley:1973,virk:1975,hinch:1977,degennes:1986,sreenivasan:2000,benzi:2005,liberzon:2006,peters:2007}. On the other hand, the effect of polymers on the bulk turbulence, away from boundaries, has received less attention. Tabor \& De Gennes~\cite{tabor:1986,degennes:1986} and Lebedev \& co-workers~\cite{balkovsky:2000,balkovsky:2001,fouxon:2003} proposed theories for homogeneous isotropic turbulence in dilute polymer solutions. 
Experiments~\cite{mccomb:1977,vandoorn:1999, liberzon:2006}
 and numerical simulations~\cite{deangelis:2005,vaithi:2006} 
have provided evidences that in bulk turbulence the small scales Eulerian quantities are damped by the polymer additives.
Validation of these theories, however, has been very difficult due to the lack of detailed experimental data. 

Recently, there has been growing interest in studying turbulence from a Lagrangian viewpoint, i.e., by following the motion of the fluid particles~\cite{falkovich:2001}. It has been shown, in particular, that the dynamics of turbulent flows with polymer additives is closely related to the correlation of velocity gradient following Lagrangian trajectories~\cite{vincenzi:2007}. 

In this Letter, we report experimental measurements of fluid acceleration and its Lagrangian correlations in intense turbulent flows of dilute polymer solutions with various concentrations. The Taylor microscale Reynolds number of the turbulent flows before adding polymers was in the range of $140 \leq R_\lambda \leq 485$. We measured fluid accelerations by optically following tracer particles with a high-speed three-dimensional particle tracking system. 
We observed strong decreases of turbulent Lagrangian accelerations and an increase in acceleration anisotropy in these flows, similar effects on Eulerian small scale quantities have been reported previously ~\cite{mccomb:1977,vandoorn:1999}.
Moreover, we observed that both effects increase with polymer concentration. 
The probability density function (PDF) of acceleration normalized by the standard deviation, however, is not affected by the polymer additives. 
The measured auto-correlation time of Lagrangian acceleration components increases with polymer concentration. 
These measurements provide detailed data of the effect of polymers on small scale Lagrangian turbulence. 
In addition, we observed an increase of the correlation time of Lagrangian acceleration magnitude, which suggests that larger scales of turbulence are also affected.

We studied a turbulent water flow generated between two counter-rotating coaxial baffled disks with blades that force the flow inertially rather than through a boundary layer, as shown in earlier experiments in a similar apparatus~\cite{cadot:1998}. 
The flow was seeded with $50 \mu$m  or $ 26\mu$m neutrally buoyant polystyrene particles.
These particles have previously been shown to faithfully follow the turbulent flow in pure water at Reynolds numbers up to $O(10^3)$~\cite{voth:2002}. 
In this study, no difference was seen between measurements made with the two different sized particles in the water or in the polymer solutions. 
In addition, within the range of polymer concentrations studied in this work (0-10ppm), we observed only small changes in fluctuation velocity with polymer concentration, comparable to the estimated uncertainties of velocity measurement.

A 35W pulsed frequency-doubled Nd:YAG laser was used to illuminate the tracer particles.
A $4.1 \times 4.1 \times 2.05$ $\mathrm{mm}^3$ volume was imaged onto four silicon strip detectors, each of which measured one coordinate of the particle position with spatial resolution $8\mu\mathrm{m}$/pixel. The silicon strip detectors can record up to 70,000 frames per second. The recording frequencies were chosen such that the time resolution of the measurement is at least 100 frames per $\tau_\eta$, sufficient for accurate acceleration measurement.
The details of the flow and imaging system have been described in~\cite{laporta:2001,voth:2002}.

The measured particle positions were connected to form three dimensional particle trajectories using the algorithms described in Ref.~\cite{voth:2002}.  
Velocities and accelerations were computed by convolution of the trajectories with an appropriate kernel based on successive derivatives of a Guassian filter of width $\sigma_k$, which performs simultaneously a low-pass filtering and the differentiation~\cite{mordant:2004}.
The variances of acceleration components, $\avar$,  for both water and the polymer solutions, were estimated as in~\cite{voth:2002}, i.e.,  $\avar$ is computed as a function of $\sigma_k$ and then is extrapolated to $\sigma_k $ going to zero, in order to remove the effect of measurement noise on acceleration variance.

We developed a protocol to prepare the polymer solutions such that the experiments are robust and reproducible. 
First a high concentration stock solution was prepared by mixing the dry polymer powder with water. 
A magnetic stir was used to mix the solution at slow speed for 16 hours.
Then the stock solution stood for several hours and was depressurized to remove  soluble gases. Finally it was gently added to the turbulence chamber to form a dilute solution at the desired concentration. We typically waited 10 to 15 minutes before starting the measurement so that the solution was well-mixed. 
In this Letter, we present measurements on solutions of polyacrylamide (PAM) 18522 from Polysciences. 
All initial stock solutions were 400 ppm except  one experiment with a final concentration of 1ppm, which had a stock solution of only 20 ppm.

According to the manufacturer,
the weight average molecular weight, $M_w$,  for the 18522 PAM is $M_w = 18 \times 10^6$amu.
The manufacturer also gave a measure of the polydispersity of the sample $M_w / M_n = 3 - 4$, where $M_n$ is the number average molecular weight.
Using the Zimm model~\cite{bird:1987}  we calculated that 
the radius of gyration $R_g = 0.5 {\mu}\mathrm{m}$, the maximum extension $R_{\max} = 77 {\mu}\mathrm{m}$ and the polymer relaxation time  $\tau_p = 43 \mathrm{ms} $ for this polymer.
As it is widely accepted, the most important dimensionless number that characterizes the polymer dynamics in turbulence is the Weissenberg number $Wi \equiv \tau_p / \tau_{\eta} $, which measures the ratio of the polymer relaxation time to the Kolmogorov time $\tau_\eta$, the smallest time scale of turbulence. Only when $Wi \gtrsim 1$, the polymer molecules can be stretched by the turbulent flow and may affect the turbulence.
The corresponding Weissenberg number for each Reynolds number that we have measured is shown in Table~\ref{tab:exp}.  Some experiments have shown that the effect of polymer on turbulence is sensitive to the polydispersity of the polymer sample with the longer polymers in the sample having a greater effect on the turbulence~\cite{berman:1977}. In view of this,  the $\tau_p$ and $Wi$ calculated from the $M_w$ may be treated as a lower bound.

\begin{table}
\begin{ruledtabular}
\begin{tabular}{cccccc}
$f_p$ & $R_\lambda$ & $\eta$ & $\tau_\eta$ & $N_f$ & $Wi$\\
(Hz) & & ($\mu$m) & (ms) & (frames/$\tau_\eta$) & \\ \hline 
 0.15 & 140 & 322 & 105 & 699 & 0.4 \\
 0.3   & 200 & 191 & 37 & 373 & 1.2 \\
 0.6   & 285 & 114 & 13.1& 175 & 3.3 \\
1.75 &  485 & 51.0 & 2.63 & 102 & 16 \\
\end{tabular}
\end{ruledtabular}
\caption{Parameters of the experiments. $f_p$ is the rotation speed of the baffled disks. $R_\lambda$ is the Taylor microscale Reynolds number. $\eta$ and $\tau_\eta$ are the Kolmogorov length and time scales, respectively. $R_\lambda$, $\eta$, and $\tau_\eta$ refer to the turbulent water flows before adding polymer solutions, and are measured as in Ref.~\cite{voth:2002}. $N_f$ is the temporal resolution of the recording system, in frames per $\tau_\eta$. $Wi \equiv \tau_p /\tau_\eta$ is the Weissenberg number.}
\label{tab:exp}
\end{table}

\begin{figure}
\centering
\includegraphics[scale=0.22]{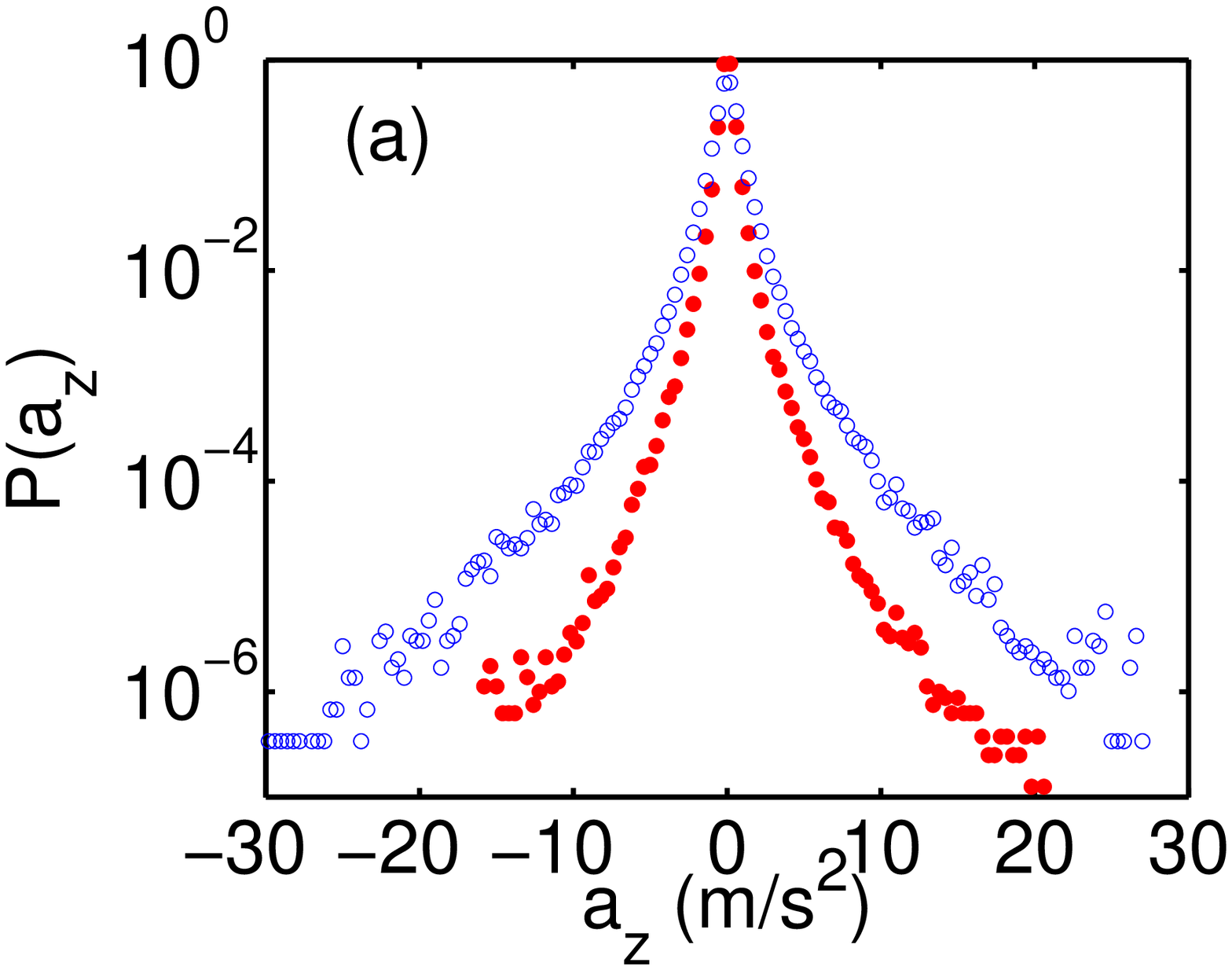}
\includegraphics[scale=0.22]{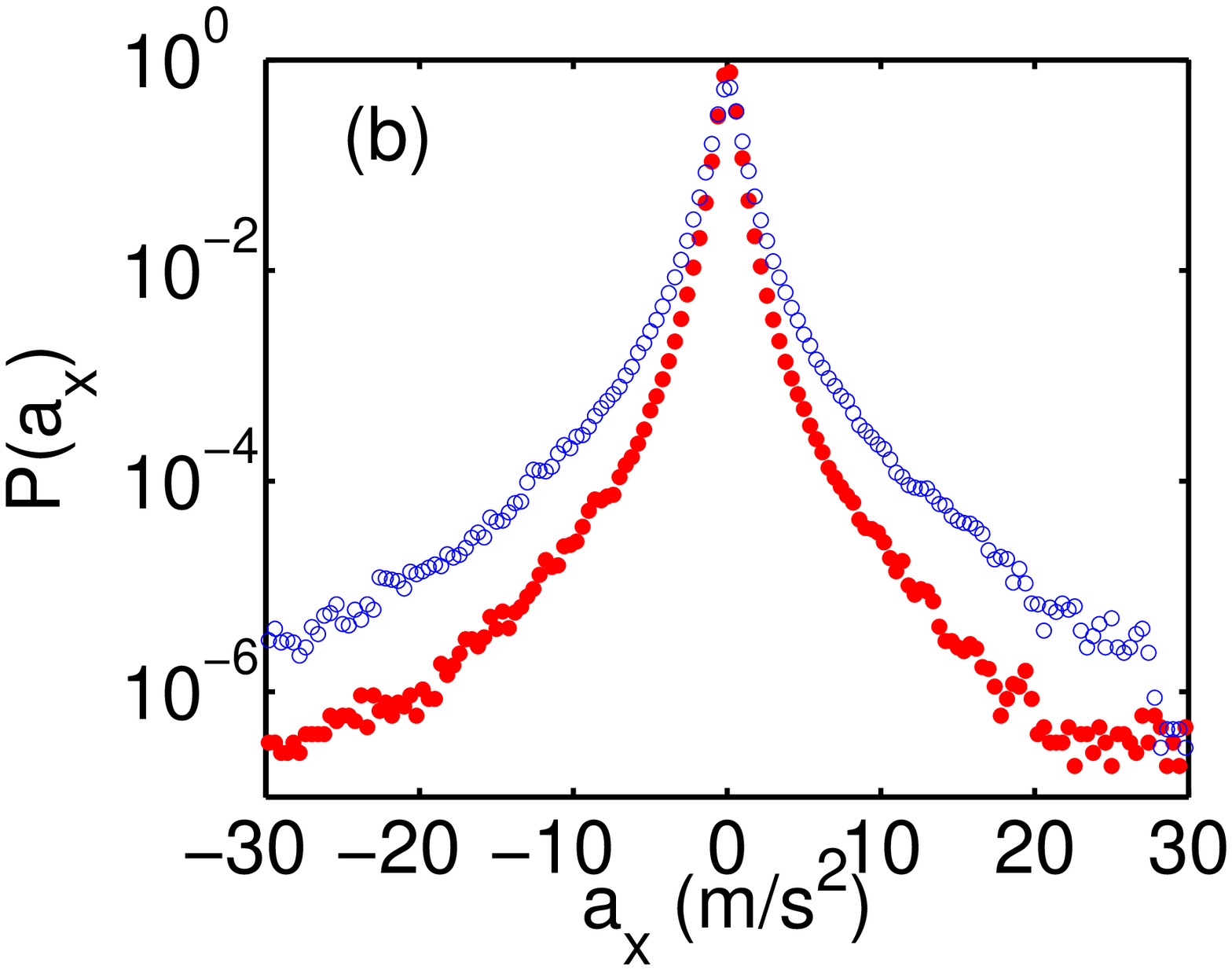}
\\
\includegraphics[scale=0.22]{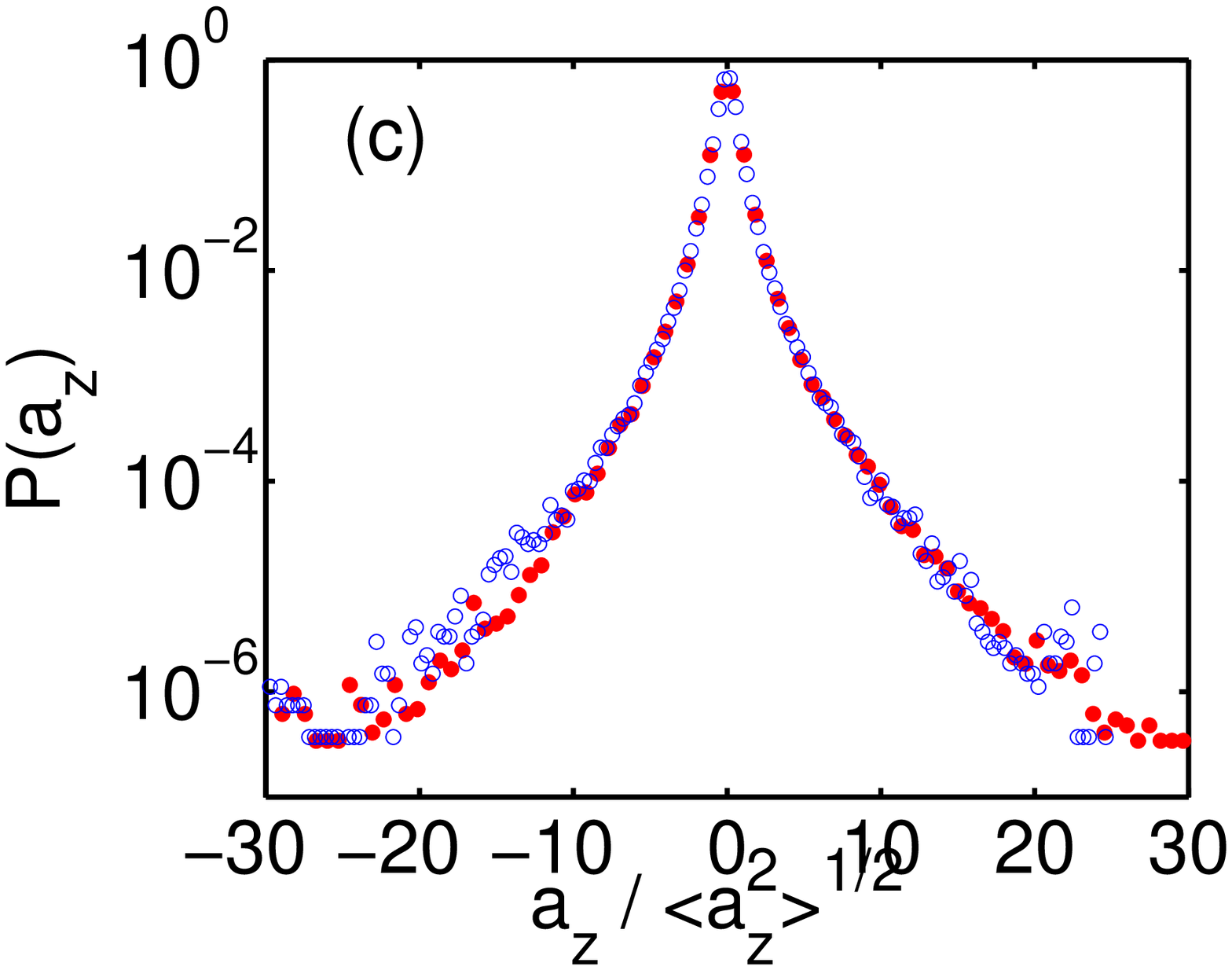}
\includegraphics[scale=0.22]{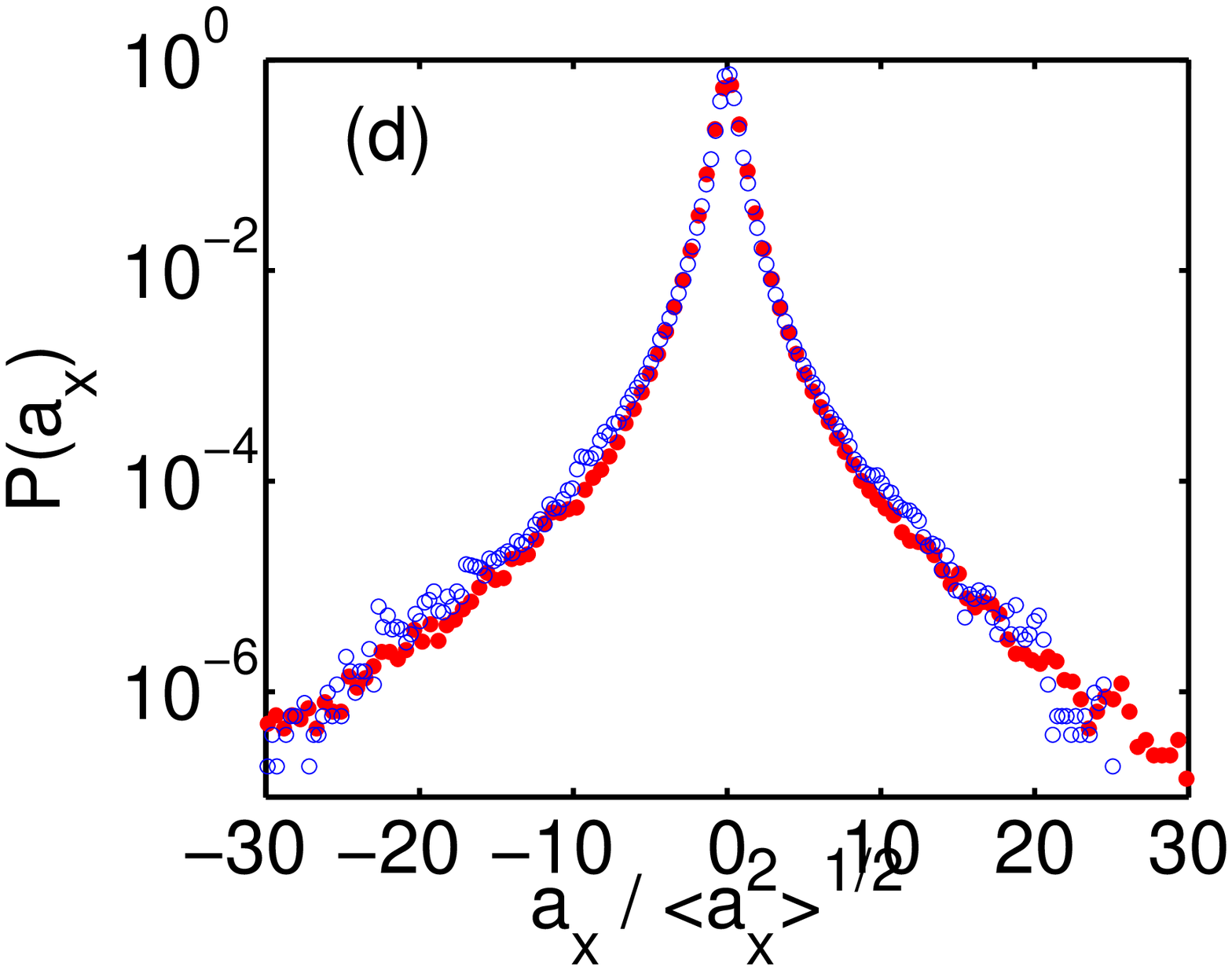}
\caption{(color online). \pdf of acceleration components at $R_\lambda = 285$.
$\circ$ are for water. $\bullet$ are for a 5ppm polymer solution. 
The width of the filter used to calculate acceleration is $\fw = 0.23 \tau_\eta$. 
(a) $P(a_z)$, (b) $P(a_x)$, (c) $P(a_x / \langle a_x^2 \rangle ^{1/2})$
(d) $P(a_z / \langle a_z^2 \rangle ^{1/2})$.
}
\label{fig:adist_pam}
\end{figure}

Figure~\ref{fig:adist_pam} shows the component acceleration \pdf for water and the polymer solution of 5ppm by weight at $R_\lambda = 285$.
The unnormalized  PDFs for the dilute polymer solution are strongly suppressed, which indicates damping of dissipation scale fluctuations by the presence of polymers.
It is interesting to note that the \pdf's of the normalized acceleration, $a_i / \langle a_i^2 \rangle^{1/2}$, 
for water and for the dilute polymer solution overlap with each other, as shown in Figures~\ref{fig:adist_pam} (c) and (d),
which suggests that the accelerations are suppressed only by a scaling factor.

\begin{figure}
\includegraphics[width=0.45\textwidth]{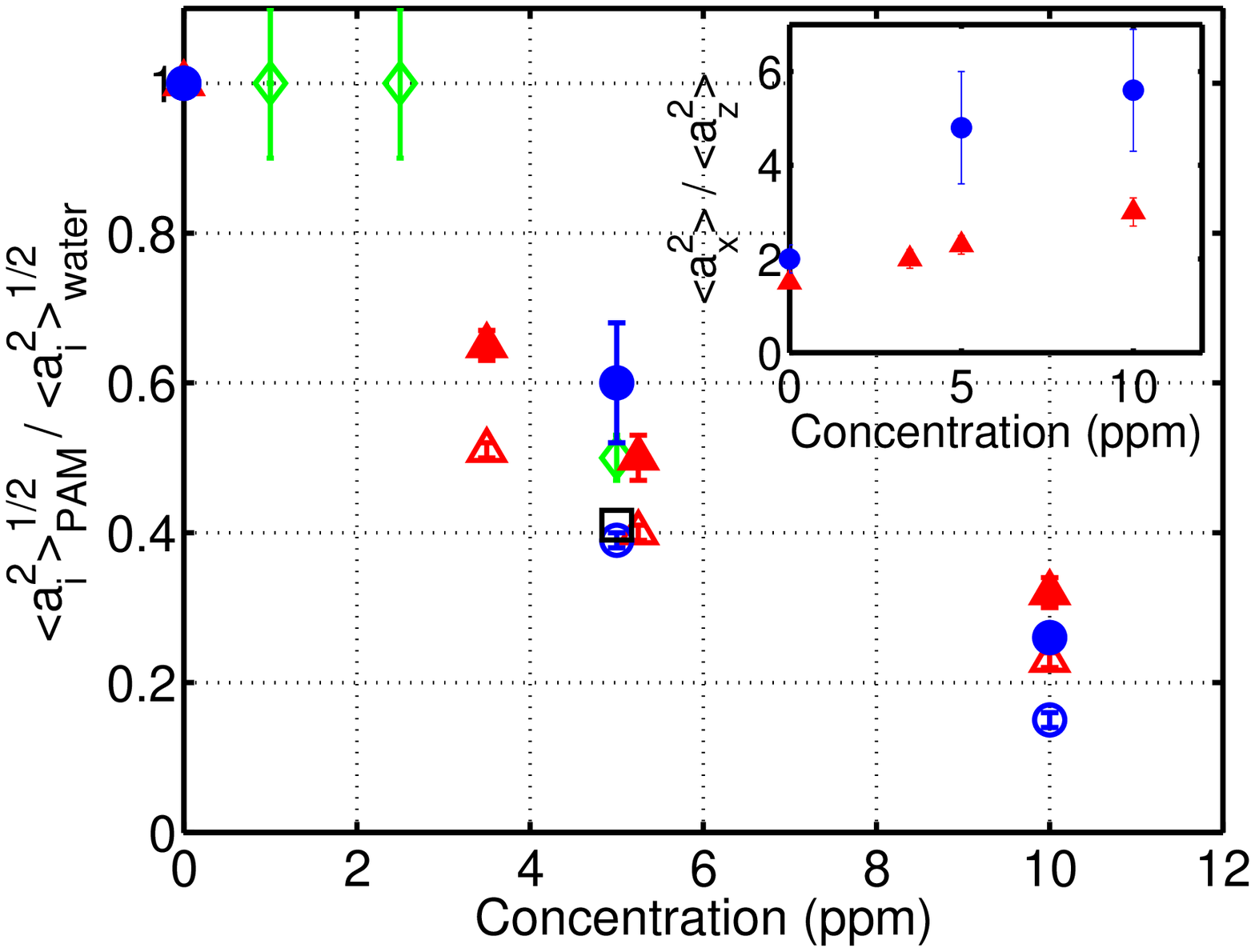}
\caption{(color online). Ratio of the \rms acceleration in polymer solution to that of water vs.
concentration of the polymer solution.
$\square$, $R_\lambda = 140$, \axial component;
$\circ$, $R_\lambda = 200$, \axial component;
$\bullet$, $R_\lambda = 200$, \transverse component;
$\triangle$, $R_\lambda = 285$, \axial component;
$\blacktriangle$, $R_\lambda = 285$, \transverse component;
$\diamond$, $R_\lambda = 485$,  \axial component; 
For the data point at $R_\lambda = 485$ and concentration 5ppm, statistics from the first 3.3 hours
of the run were used. The variance changes less than 10\% over this time.
Data points at $f_p = 0.6$ hz and concentration 5 ppm were shifted to 5.25 ppm for clarity.
INSET: Ratio of the \transverse acceleration variance to the \axial acceleration variance 
as a function of concentration of the polymer solution.
$\bullet$, $R_\lambda = 200$;
$\blacktriangle$, $R_\lambda = 285$.
}
\label{fig:ratios}
\end{figure}

The rms acceleration, $\avar^{1/2}$,  in the polymer solutions is dramatically lower than in water.
Figure~\ref{fig:ratios} shows the ratio of the  \rms acceleration in polymer solutions to that of water
for the same rotation rate of the propellers. 
In the range of polymer concentrations investigated, the dependence on concentration is roughly linear. 
There is no clear dependence on $R_\lambda$ or $Wi$. 

It should be pointed out that the decrease in acceleration variance can not be explained by a decrease in energy dissipation rate $\varepsilon$. In earlier experiments carried out in similar geometry, it has been shown that the total energy dissipation rate remains unchanged when polymers were injected into the flow if baffled disks were used to provide the driving~\cite{cadot:1998}. The inertial forcing by the vanes was not altered by the presence of polymers. 

The anisotropy of the small scales is also increasing upon addition of the polymer. 
The insert in Figure~\ref{fig:ratios} shows the ratio of the \transverse acceleration variance to
the \axial acceleration variance as a function of polymer concentration. 
If the polymers were causing the energy cascade to be truncated at a smaller wave number, then the separation between large and small scales would be reduced and the anisotropy of the large scales would affect the small scales more.
This is similar to seeing the anisotropy increase with decreasing $\Rl$. 
 The increase of anisotropy might also come from the alignment of polymers in particular directions due to the large scale anisotropic flow. Since the polymer responses in longitudinal and transverse directions (relative to the alignment of polymers) are different~\cite{hatfield:1999}, this difference could appear as anisotropy in acceleration.

Since the dynamics of polymer solutions depends on the velocity field following Lagrangian trajectories, we studied the Lagrangian acceleration autocovariance $c_{ij}(\tau) = {\left\langle a_i(t) a_j(t+\tau) \right\rangle}$
and the Lagrangian autocorrelation $C_{ij}(\tau) = \frac{\left\langle a_i(t) a_j(t+\tau) \right\rangle} {
\left(\left\langle a_i(t)^2\right\rangle
\left\langle a_j(t+\tau)^2\right\rangle\right)^{1/2}
} . $
To account for the effect of the finite volume bias, we calculated the modified autocovariance $\tilde{c}_{ij}(\tau)$
\begin{equation}
\tilde{c}_{ij}(\tau) = \frac{\left\langle a_i(t) a_i(t+\tau) \right\rangle_{t\&\tau}}
{
\left(\left\langle a_i(t)^2\right\rangle_{t\&\tau}
\left\langle a_i(t+\tau)^2\right\rangle_{t\&\tau}\right)^{1/2} }
\left\langle a_i(t)^2\right\rangle_{t} ,
\label{eq:auto2} 
\end{equation}
in which $\langle \rangle_{t\&\tau}$ indicates
that only measurements of acceleration where a measurement at both $t$ and $t+\tau$ existed 
on a  particle track were averaged over and $\langle \rangle_{t}$ indicates that all measurements of acceleration 
were averaged over. 
The autocorrelation is then
\begin{equation}
%\tilde{C}_{ij}(\tau) = \frac{\tilde{c}_{ij}(\tau) }{a_0 \nu^{-1/2} \epsilon^{3/2}}
\tilde{C}_{ij}(\tau) = \frac{\tilde{c}_{ij}(\tau) }{\avar^{1/2}} ,
\label{eq:acor2} 
\end{equation}
where $\avar$  is determined from the procedure described above.

\begin{figure}
\includegraphics[width=0.45\textwidth]{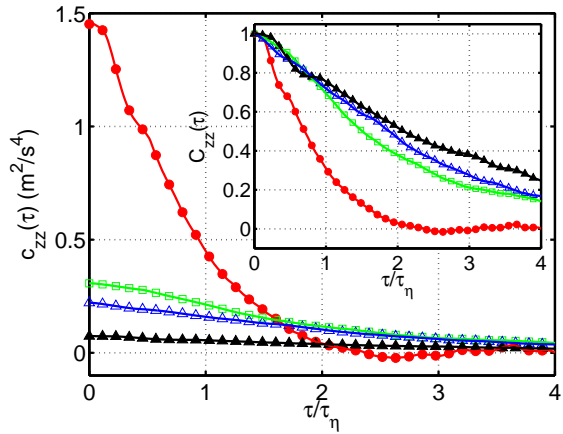}
\caption{(color online). Autocovariance of the \axial acceleration component
for $R_\lambda = 285$ at different polymer concentrations.
$\bullet$ water;
$\blacktriangle$ 10ppm;
$\triangle$ 5ppm;
$\square$ 3.5ppm.
INSET: Corresponding autocorrelations.
}
\label{fig:auto_06_pam}
\end{figure}

Figure~\ref{fig:auto_06_pam} shows $\tilde{c}_{ij}(\tau)$ and  $\tilde{C}_{ij}(\tau)$ for water and polymer.
The height of the autocovariance is depressed because, as we have discussed above,
$\avar$ is dramatically decreased in the polymer solutions. 
It has been shown that the integral of the acceleration correlation vanishes~\cite{tennekes:1972} and the zero-crossing time of the correlation curve is usually used as a measure of the correlation time~\cite{yeung:1989, mordant:2004c}. The curve at the zero-crossing has a shallow slope.  Small measurement errors in the acceleration correlation can cause large deviations in the zero-crossing time. A more robust measure is  the exponential decay time  $\tau_{\ooe}$, defined such that $\tilde{C}(\tau_{\ooe}) = 1/e$.
In pure water, the $\tau_\ooe$ times are approximately 0.7-0.8 times the Kolmogorov times.
As shown in Figure~\ref{fig:auto_06_pam}, in the range of polymer concentration used, the $\tau_\ooe$ for polymer is about 3-4 times the $\tau_\ooe$ for water and increases with polymer concentration.

The width of the acceleration correlation is proportional to the Kolmogorov time, $\tau_\eta$. 
In pure water we observe that component acceleration correlation functions for $R_\lambda = 285$ to 970 
collapse when the time is normalized by $\tau_\eta$~\cite{crawford:2004}.
The Kolmogorov time in turn is proportional to $\nu^{1/2}$. 
The acceleration variance  is proportional to $\nu^{-1/2}$.
If the polymer solutions were to behave simply as if they had an increased effective viscosity,  then
we would expect  the decrease in the acceleration variance to match the increase in
the correlation time. Instead we find that  the decrease in $\avar$ is much larger than the increase in the 
autocorrelation time. 
The acceleration
variance for a 3.5 ppm solution is more than 4 times higher than for a 10 ppm solution.
However the $\tau_\ooe$ for the 10 ppm solutions is not even twice as much as that of the 3.5 ppm solution.
The backreaction of the polymer is obviously more complex than just a change in the bulk viscosity and is most likely anisotropic and dependent on the local velocity gradients~\cite{vincenzi:2007} as for example the elongational viscosity is known to change by orders of magnitude but not the shear viscosity~\cite{hinch:1977}.

\begin{figure}
\centering
\includegraphics[width = 0.45\textwidth]{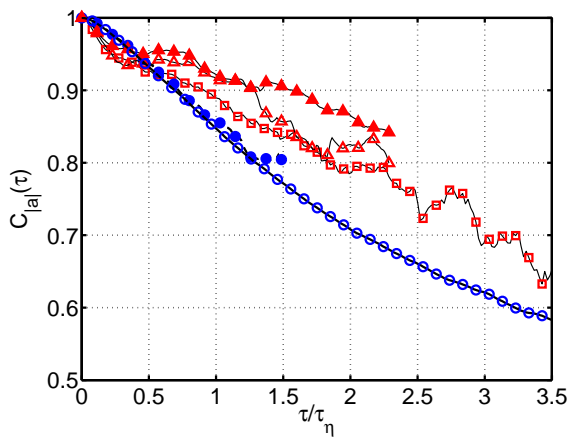}
\caption{(color online). $C_{|a|}(\tau)$ for water and polymer. 
$\circ$ water $R_\lambda = 485$, $\fw = 0.35 \tau_\eta $.
$\bullet$ water $R_\lambda = 285$, $\fw = 0.31 \tau_\eta$.
$\square$ 5ppm, $R_\lambda = 485$, $\fw = 0.35 \tau_\eta$.
$\triangle$ 5ppm, $R_\lambda = 285$, $\fw =  0.31 \tau_\eta$.
$\blacktriangle$ 10ppm, $R_\lambda = 285$, $\fw =  0.31 \tau_\eta $.
The polymer statistics for $R_\lambda = 485$ were taken from the first 3.3 hours of the run.
}
\label{fig:amag_cor_pam}
\end{figure}

We also measured the autocorrelation of the magnitude of Lagrangian acceleration, defined as 
$C_{| \mathbf a |}=\frac{\langle (| \mathbf a (t)|-\langle | \mathbf a | \rangle)(| \mathbf a (t+\tau) |-\langle |\mathbf a |\rangle)\rangle}{\langle (| \mathbf a (t)|-\langle | \mathbf a | \rangle)^2\rangle}$
and  obtained using an estimator equivalent to Eq.~\eqref{eq:acor2}.
The results are shown in Fig.~\ref{fig:amag_cor_pam}.
The time scale given by the acceleration magnitude correlation increases with polymer concentration.
This may indicate that the large scales of the flow are affected by the polymer as well, 
since the acceleration magnitude is correlated over much longer time scales of the flow
than the acceleration components.
Due to the finite size of the measurement volume, we only measured the correlation of the acceleration magnitude over a few $\tau_\eta$.
It is possible that, at longer times, the curve for the polymer solution 
rejoins that for water. 

In summary, we studied the effect of polymers on the small scales of turbulence by measuring the fluid acceleration in turbulent flows of dilute polymer solutions. We observed strong decreases of acceleration variance in polymer solutions while the shape of the acceleration PDF remains the same as in water. The effect on acceleration variance was insensitive to the Reynolds number or the Weissenberg number. The Lagrangian correlation time of acceleration components increases with polymer concentration. The observed effect of polymer additives cannot be explained by a change of viscosity alone. In addition, the measured Lagrangian correlation time of acceleration magnitude increases with concentration, which implies that polymers may affect larger scales. 

This research was supported by the
National Science Foundation under grants PHY-9988755 and PHY-0216406.

%\bibliographystyle{apsrev}
%\bibliography{turb_refs.bib}

\begin{thebibliography}{30}
\expandafter\ifx\csname natexlab\endcsname\relax\def\natexlab#1{#1}\fi
\expandafter\ifx\csname bibnamefont\endcsname\relax
  \def\bibnamefont#1{#1}\fi
\expandafter\ifx\csname bibfnamefont\endcsname\relax
  \def\bibfnamefont#1{#1}\fi
\expandafter\ifx\csname citenamefont\endcsname\relax
  \def\citenamefont#1{#1}\fi
\expandafter\ifx\csname url\endcsname\relax
  \def\url#1{\texttt{#1}}\fi
\expandafter\ifx\csname urlprefix\endcsname\relax\def\urlprefix{URL }\fi
\providecommand{\bibinfo}[2]{#2}
\providecommand{\eprint}[2][]{\url{#2}}

\bibitem[{\citenamefont{Toms}(1948)}]{toms:1948}
\bibinfo{author}{\bibfnamefont{B.~A.} \bibnamefont{Toms}},
  \bibinfo{journal}{Proc.~1st Int.~Rheol.~Congr.} \textbf{\bibinfo{volume}{2}},
  \bibinfo{pages}{135} (\bibinfo{year}{1948}).

\bibitem[{\citenamefont{Lumley}(1973)}]{lumley:1973}
\bibinfo{author}{\bibfnamefont{J.~L.} \bibnamefont{Lumley}},
  \bibinfo{journal}{J.~Polymer Sci.} \textbf{\bibinfo{volume}{7}},
  \bibinfo{pages}{263} (\bibinfo{year}{1973}).

\bibitem[{\citenamefont{Virk}(1975)}]{virk:1975}
\bibinfo{author}{\bibfnamefont{P.~S.} \bibnamefont{Virk}},
  \bibinfo{journal}{AIChE J.} \textbf{\bibinfo{volume}{21}},
  \bibinfo{pages}{625} (\bibinfo{year}{1975}).

\bibitem[{\citenamefont{De~Gennes}(1986)}]{degennes:1986}
\bibinfo{author}{\bibfnamefont{P.~G.} \bibnamefont{De~Gennes}},
  \bibinfo{journal}{Physica A} \textbf{\bibinfo{volume}{140}},
  \bibinfo{pages}{9} (\bibinfo{year}{1986}).

\bibitem[{\citenamefont{Sreenivasan and White}(2000)}]{sreenivasan:2000}
\bibinfo{author}{\bibfnamefont{K.~R.} \bibnamefont{Sreenivasan}}
  \bibnamefont{and} \bibinfo{author}{\bibfnamefont{C.~M.} \bibnamefont{White}},
  \bibinfo{journal}{J.~Fluid Mech.} \textbf{\bibinfo{volume}{409}},
  \bibinfo{pages}{149} (\bibinfo{year}{2000}).

\bibitem[{\citenamefont{Benzi et~al.}(2005)\citenamefont{Benzi, De~Angelis,
  L'vov, and Procaccia}}]{benzi:2005}
\bibinfo{author}{\bibfnamefont{R.}~\bibnamefont{Benzi}},
  \bibinfo{author}{\bibfnamefont{E.}~\bibnamefont{De~Angelis}},
  \bibinfo{author}{\bibfnamefont{V.~S.} \bibnamefont{L'vov}}, \bibnamefont{and}
  \bibinfo{author}{\bibfnamefont{I.}~\bibnamefont{Procaccia}},
  \bibinfo{journal}{Phys.~Rev.~Lett.} \textbf{\bibinfo{volume}{95}},
  \bibinfo{pages}{194502} (\bibinfo{year}{2005}).

\bibitem[{\citenamefont{Liberzon et~al.}(2006)\citenamefont{Liberzon, Guala,
  Kinzelbach, and Tsinober}}]{liberzon:2006}
\bibinfo{author}{\bibfnamefont{A.}~\bibnamefont{Liberzon}},
  \bibinfo{author}{\bibfnamefont{M.}~\bibnamefont{Guala}},
  \bibinfo{author}{\bibfnamefont{W.}~\bibnamefont{Kinzelbach}},
  \bibnamefont{and} \bibinfo{author}{\bibfnamefont{A.}~\bibnamefont{Tsinober}},
  \bibinfo{journal}{Phys.~Fluids} \textbf{\bibinfo{volume}{18}},
  \bibinfo{pages}{125101} (\bibinfo{year}{2006}).

\bibitem[{\citenamefont{Peters and Schumacher}(2007)}]{peters:2007}
\bibinfo{author}{\bibfnamefont{T.}~\bibnamefont{Peters}} \bibnamefont{and}
  \bibinfo{author}{\bibfnamefont{J.}~\bibnamefont{Schumacher}},
  \bibinfo{journal}{Phys.~Fluids} \textbf{\bibinfo{volume}{19}},
  \bibinfo{pages}{065109} (\bibinfo{year}{2007}).

\bibitem[{\citenamefont{Hinch}(1977)}]{hinch:1977}
\bibinfo{author}{\bibfnamefont{E.~J.} \bibnamefont{Hinch}},
  \bibinfo{journal}{Phys.~Fluids} \textbf{\bibinfo{volume}{20}},
  \bibinfo{pages}{S22} (\bibinfo{year}{1977}).

\bibitem[{\citenamefont{Tabor and De~Gennes}(1986)}]{tabor:1986}
\bibinfo{author}{\bibfnamefont{M.}~\bibnamefont{Tabor}} \bibnamefont{and}
  \bibinfo{author}{\bibfnamefont{P.~G.} \bibnamefont{De~Gennes}},
  \bibinfo{journal}{Europhys.~Lett.} \textbf{\bibinfo{volume}{2}},
  \bibinfo{pages}{519} (\bibinfo{year}{1986}).

\bibitem[{\citenamefont{Balkovsky et~al.}(2000)\citenamefont{Balkovsky, Fouxon,
  and Lebedev}}]{balkovsky:2000}
\bibinfo{author}{\bibfnamefont{E.}~\bibnamefont{Balkovsky}},
  \bibinfo{author}{\bibfnamefont{A.}~\bibnamefont{Fouxon}}, \bibnamefont{and}
  \bibinfo{author}{\bibfnamefont{V.}~\bibnamefont{Lebedev}},
  \bibinfo{journal}{Phys.~Rev.~Lett.} \textbf{\bibinfo{volume}{84}},
  \bibinfo{pages}{4765} (\bibinfo{year}{2000}).

\bibitem[{\citenamefont{Balkovsky et~al.}(2001)\citenamefont{Balkovsky, Fouxon,
  and Lebedev}}]{balkovsky:2001}
\bibinfo{author}{\bibfnamefont{E.}~\bibnamefont{Balkovsky}},
  \bibinfo{author}{\bibfnamefont{A.}~\bibnamefont{Fouxon}}, \bibnamefont{and}
  \bibinfo{author}{\bibfnamefont{V.}~\bibnamefont{Lebedev}},
  \bibinfo{journal}{Phys.~Rev.~E} \textbf{\bibinfo{volume}{64}},
  \bibinfo{pages}{056301} (\bibinfo{year}{2001}).

\bibitem[{\citenamefont{Fouxon and Lebedev}(2003)}]{fouxon:2003}
\bibinfo{author}{\bibfnamefont{A.}~\bibnamefont{Fouxon}} \bibnamefont{and}
  \bibinfo{author}{\bibfnamefont{V.}~\bibnamefont{Lebedev}},
  \bibinfo{journal}{Phys.~Fluids} \textbf{\bibinfo{volume}{15}},
  \bibinfo{pages}{2060} (\bibinfo{year}{2003}).

\bibitem[{\citenamefont{McComb et~al.}(1977)\citenamefont{McComb, Allan, and
  Greated}}]{mccomb:1977}
\bibinfo{author}{\bibfnamefont{W.~D.} \bibnamefont{McComb}},
  \bibinfo{author}{\bibfnamefont{J.}~\bibnamefont{Allan}}, \bibnamefont{and}
  \bibinfo{author}{\bibfnamefont{C.~A.} \bibnamefont{Greated}},
  \bibinfo{journal}{Phys.~Fluids} \textbf{\bibinfo{volume}{20}},
  \bibinfo{pages}{873} (\bibinfo{year}{1977}).

\bibitem[{\citenamefont{van Doorn et~al.}(1999)\citenamefont{van Doorn, White,
  and Sreenivasan}}]{vandoorn:1999}
\bibinfo{author}{\bibfnamefont{E.}~\bibnamefont{van Doorn}},
  \bibinfo{author}{\bibfnamefont{C.~M.} \bibnamefont{White}}, \bibnamefont{and}
  \bibinfo{author}{\bibfnamefont{K.~R.} \bibnamefont{Sreenivasan}},
  \bibinfo{journal}{Phys.~Fluids} \textbf{\bibinfo{volume}{11}},
  \bibinfo{pages}{2387} (\bibinfo{year}{1999}).

\bibitem[{\citenamefont{De~Angelis et~al.}(2005)\citenamefont{De~Angelis,
  Casciola, Benzi, and Piva}}]{deangelis:2005}
\bibinfo{author}{\bibfnamefont{E.}~\bibnamefont{De~Angelis}},
  \bibinfo{author}{\bibfnamefont{C.~M.} \bibnamefont{Casciola}},
  \bibinfo{author}{\bibfnamefont{R.}~\bibnamefont{Benzi}}, \bibnamefont{and}
  \bibinfo{author}{\bibfnamefont{R.}~\bibnamefont{Piva}},
  \bibinfo{journal}{J.~Fluid Mech.} \textbf{\bibinfo{volume}{531}},
  \bibinfo{pages}{1} (\bibinfo{year}{2005}).

\bibitem[{\citenamefont{Vaithianatha et~al.}(2006)\citenamefont{Vaithianatha,
  Robert, Brasseur, and Collins}}]{vaithi:2006}
\bibinfo{author}{\bibfnamefont{T.}~\bibnamefont{Vaithianatha}},
  \bibinfo{author}{\bibfnamefont{A.}~\bibnamefont{Robert}},
  \bibinfo{author}{\bibfnamefont{J.~G.} \bibnamefont{Brasseur}},
  \bibnamefont{and} \bibinfo{author}{\bibfnamefont{L.~R.}
  \bibnamefont{Collins}}, \bibinfo{journal}{J.~Non-Newtonian Fluid Mech.}
  \textbf{\bibinfo{volume}{140}}, \bibinfo{pages}{3} (\bibinfo{year}{2006}).

\bibitem[{\citenamefont{Falkovich et~al.}(2001)\citenamefont{Falkovich,
  Gawedzki, and Vergassola}}]{falkovich:2001}
\bibinfo{author}{\bibfnamefont{G.}~\bibnamefont{Falkovich}},
  \bibinfo{author}{\bibfnamefont{K.}~\bibnamefont{Gawedzki}}, \bibnamefont{and}
  \bibinfo{author}{\bibfnamefont{M.}~\bibnamefont{Vergassola}},
  \bibinfo{journal}{Rev.~Mod.~Phys.} \textbf{\bibinfo{volume}{73}},
  \bibinfo{pages}{913} (\bibinfo{year}{2001}).

\bibitem[{\citenamefont{Vincenzi et~al.}(2007)\citenamefont{Vincenzi, Jin,
  Bodenschatz, and Collins}}]{vincenzi:2007}
\bibinfo{author}{\bibfnamefont{D.}~\bibnamefont{Vincenzi}},
  \bibinfo{author}{\bibfnamefont{S.}~\bibnamefont{Jin}},
  \bibinfo{author}{\bibfnamefont{E.}~\bibnamefont{Bodenschatz}},
  \bibnamefont{and} \bibinfo{author}{\bibfnamefont{L.~R.}
  \bibnamefont{Collins}}, \bibinfo{journal}{Phys.~Rev.~Lett.}
  \textbf{\bibinfo{volume}{98}}, \bibinfo{pages}{024503}
  (\bibinfo{year}{2007}).

\bibitem[{\citenamefont{Cadot et~al.}(1998)\citenamefont{Cadot, Bonn, and
  Douady}}]{cadot:1998}
\bibinfo{author}{\bibfnamefont{O.}~\bibnamefont{Cadot}},
  \bibinfo{author}{\bibfnamefont{D.}~\bibnamefont{Bonn}}, \bibnamefont{and}
  \bibinfo{author}{\bibfnamefont{S.}~\bibnamefont{Douady}},
  \bibinfo{journal}{Phys.~Fluids} \textbf{\bibinfo{volume}{10}},
  \bibinfo{pages}{426} (\bibinfo{year}{1998}).

\bibitem[{\citenamefont{Voth et~al.}(2002)\citenamefont{Voth, La~Porta,
  Crawford, Alexander, and Bodenschatz}}]{voth:2002}
\bibinfo{author}{\bibfnamefont{G.~A.} \bibnamefont{Voth}},
  \bibinfo{author}{\bibfnamefont{A.}~\bibnamefont{La~Porta}},
  \bibinfo{author}{\bibfnamefont{A.~M.} \bibnamefont{Crawford}},
  \bibinfo{author}{\bibfnamefont{J.}~\bibnamefont{Alexander}},
  \bibnamefont{and}
  \bibinfo{author}{\bibfnamefont{E.}~\bibnamefont{Bodenschatz}},
  \bibinfo{journal}{J.~Fluid Mech.} \textbf{\bibinfo{volume}{469}},
  \bibinfo{pages}{121} (\bibinfo{year}{2002}).

\bibitem[{\citenamefont{La~Porta et~al.}(2001)\citenamefont{La~Porta, Voth,
  Crawford, Alexander, and Bodenschatz}}]{laporta:2001}
\bibinfo{author}{\bibfnamefont{A.}~\bibnamefont{La~Porta}},
  \bibinfo{author}{\bibfnamefont{G.~A.} \bibnamefont{Voth}},
  \bibinfo{author}{\bibfnamefont{A.~M.} \bibnamefont{Crawford}},
  \bibinfo{author}{\bibfnamefont{J.}~\bibnamefont{Alexander}},
  \bibnamefont{and}
  \bibinfo{author}{\bibfnamefont{E.}~\bibnamefont{Bodenschatz}},
  \bibinfo{journal}{Nature} \textbf{\bibinfo{volume}{409}},
  \bibinfo{pages}{1017} (\bibinfo{year}{2001}).

\bibitem[{\citenamefont{Mordant
  et~al.}(2004{\natexlab{a}})\citenamefont{Mordant, Crawford, and
  Bodenschatz}}]{mordant:2004}
\bibinfo{author}{\bibfnamefont{N.}~\bibnamefont{Mordant}},
  \bibinfo{author}{\bibfnamefont{A.~M.} \bibnamefont{Crawford}},
  \bibnamefont{and}
  \bibinfo{author}{\bibfnamefont{E.}~\bibnamefont{Bodenschatz}},
  \bibinfo{journal}{Physica D} \textbf{\bibinfo{volume}{193}},
  \bibinfo{pages}{245} (\bibinfo{year}{2004}{\natexlab{a}}).

\bibitem[{\citenamefont{Bird et~al.}(1987)\citenamefont{Bird, Amstrong, and
  Hassager}}]{bird:1987}
\bibinfo{author}{\bibfnamefont{R.~B.} \bibnamefont{Bird}},
  \bibinfo{author}{\bibfnamefont{R.~C.} \bibnamefont{Amstrong}},
  \bibnamefont{and} \bibinfo{author}{\bibfnamefont{S.}~\bibnamefont{Hassager}},
  \emph{\bibinfo{title}{Dynamics of Polymeric Liquids}}
  (\bibinfo{publisher}{John Wiley and Sons}, \bibinfo{address}{New York},
  \bibinfo{year}{1987}).

\bibitem[{\citenamefont{Berman}(1977)}]{berman:1977}
\bibinfo{author}{\bibfnamefont{N.~S.} \bibnamefont{Berman}},
  \bibinfo{journal}{Phys.~Fluids} \textbf{\bibinfo{volume}{20}},
  \bibinfo{pages}{715} (\bibinfo{year}{1977}).

\bibitem[{\citenamefont{Hatfield and Quake}(1999)}]{hatfield:1999}
\bibinfo{author}{\bibfnamefont{J.~W.} \bibnamefont{Hatfield}} \bibnamefont{and}
  \bibinfo{author}{\bibfnamefont{S.~R.} \bibnamefont{Quake}},
  \bibinfo{journal}{Phys.~Rev.~Lett.} \textbf{\bibinfo{volume}{82}},
  \bibinfo{pages}{3548} (\bibinfo{year}{1999}).

\bibitem[{\citenamefont{Tennekes and Lumley}(1972)}]{tennekes:1972}
\bibinfo{author}{\bibfnamefont{T.}~\bibnamefont{Tennekes}} \bibnamefont{and}
  \bibinfo{author}{\bibfnamefont{J.~L.} \bibnamefont{Lumley}},
  \emph{\bibinfo{title}{A First Course in Turbulence}} (\bibinfo{publisher}{The
  MIT Press}, \bibinfo{address}{Cambridge, USA}, \bibinfo{year}{1972}).

\bibitem[{\citenamefont{Yeung and Pope}(1989)}]{yeung:1989}
\bibinfo{author}{\bibfnamefont{P.~K.} \bibnamefont{Yeung}} \bibnamefont{and}
  \bibinfo{author}{\bibfnamefont{S.~B.} \bibnamefont{Pope}},
  \bibinfo{journal}{J.~Fluid Mech.} \textbf{\bibinfo{volume}{207}},
  \bibinfo{pages}{531} (\bibinfo{year}{1989}).

\bibitem[{\citenamefont{Mordant
  et~al.}(2004{\natexlab{b}})\citenamefont{Mordant, Crawford, and
  Bodenschatz}}]{mordant:2004c}
\bibinfo{author}{\bibfnamefont{N.}~\bibnamefont{Mordant}},
  \bibinfo{author}{\bibfnamefont{A.~M.} \bibnamefont{Crawford}},
  \bibnamefont{and}
  \bibinfo{author}{\bibfnamefont{E.}~\bibnamefont{Bodenschatz}},
  \bibinfo{journal}{Phys.~Rev.~Lett.} \textbf{\bibinfo{volume}{93}},
  \bibinfo{pages}{214501} (\bibinfo{year}{2004}{\natexlab{b}}).

\bibitem[{\citenamefont{Crawford}(2004)}]{crawford:2004}
\bibinfo{author}{\bibfnamefont{A.~M.} \bibnamefont{Crawford}}, Ph.D. thesis,
  \bibinfo{school}{Cornell University} (\bibinfo{year}{2004}).

\end{thebibliography}

\end{document}